\documentclass[preprint,10pt,5p,twocolumn,authoryear]{elsarticle}

\usepackage{lipsum}

\makeatletter 
\def\ps@pprintTitle{ 
 \let\@oddhead\@empty 
 \let\@evenhead\@empty 
 \def\@oddfoot{\footnotesize{\textit{Accepted for publication in Astronomy \& Computing\hfill\today}}} 
 \def\@evenfoot{\thepage\hfill}} 
\makeatother

\usepackage{hyperref}

\journal{Astronomy and Computing}

\usepackage{relsize}
\usepackage{amsmath,amssymb}

\begin{document}

\begin{frontmatter}

\title{E0102-VR: exploring the scientific potential of Virtual Reality for observational astrophysics}

\author{Enrico Baracaglia}
\address{Institut Sup\'erieur de l'A\'eronautique et de l'Espace (ISAE-SUPAERO), 10 Avenue Edouard Belin, 31400 Toulouse, France}

\author{Fr\'ed\'eric P.A. Vogt\corref{cor1}}
\cortext[cor1]{ESO Fellow}
\ead{frederic.vogt@alumni.anu.edu.au}
\ead[url]{http://fpavogt.github.io}

\address{European Southern Observatory, Av. Alonso de C\'ordova 3107, Vitacura, Santiago, Chile}

\begin{abstract}
Virtual Reality (VR) technology has been subject to a rapid democratization in recent years, driven in large by the entertainment industry, and epitomized by the emergence of consumer-grade, plug-and-play, room-scale VR devices. To explore the scientific potential of this technology for the field of observational astrophysics, we have created an experimental VR application: \textit{E0102-VR}. The specific scientific goal of this application is to facilitate the characterization of the 3D structure of the oxygen-rich ejecta in the young supernova remnant 1E\,0102.2-7219 in the Small Magellanic Cloud. Using \textit{E0102-VR}, we measure the physical size of two large cavities in the system, including a (7.0$\pm$0.5)\,pc-long funnel structure on the far-side of the remnant. The \textit{E0102-VR} application, albeit experimental, demonstrates the benefits of using human depth perception for a rapid and accurate characterization of complex 3D structures. Given the implementation costs (time-wise) of a dedicated VR application like \textit{E0102-VR}, we conclude that the future of VR for scientific purposes in astrophysics most likely resides in the development of a robust, generic application dedicated to the exploration and visualization of 3D observational datasets, akin to a ``ds9-VR''.
\end{abstract}


\end{frontmatter}

\section{Introduction} \label{sec:intro}

The emergence of plug-and-play Virtual Reality (VR) devices is opening new avenues for the visualization and communication of scientific datasets. The perceived maturity of VR technology is reflected in its location within the (informative, albeit highly subjective) Hype Cycle published yearly by Gartner\footnote{\url{https://www.gartner.com/en/research/methodologies/gartner-hype-cycle}}. Specifically, VR technology was deemed to have moved out of the ``trough of disillusionment'' into the ``slope of enlightenment'' in 2016, with a predicted reach of the ``plateau of productivity'' within 5-10 years. Undoubtedly, exploring the potential of VR for astrophysics is a timely venture.  

The development of consumer-grade VR devices, which can provide robust, room-scale, immersive VR experiences to the general public, has been largely driven by the entertainment industry. In early 2019, the two gaming engines \textsc{unity3d} and \textsc{unrealengine} both allow to assemble experimental VR applications for free\footnote{For the exact licensing information, see \url{https://unity3d.com/legal/terms-of-service} and \url{https://www.unrealengine.com/en-US/eula}}. This fact greatly facilitates the creation of non-commercially-viable VR experiences, such as applications for scientific outreach purposes, for example.

The potential of VR for astrophysics has already been identified, be it in the form of immersive and interactive experiences \citep[see e.g.][]{Farr2009,Djorgovski2013,Ferrand2016, Fluke2018} or 360$^{\circ}$ videos \citep{Russell2017,Davelaar2018}. For outreach purposes, VR technology can provide the general public with a highly enticing means to apprehend complex astrophysical datasets and concepts \citep{Ferrand2018, Arcand2018}. The use of VR for the purpose of scientific analysis in astrophysics is perhaps less evident, but it is certainly not an unchartered territory \citep[][; Law et al., in preparation; McGraw et al., in preparation]{Lee2018,Romano2019}. Dedicated experiments include the ``VRlab'' developed at the Universit\"ats-Sternwarte M\"unchen\footnote{\url{http://www.usm.uni-muenchen.de/vrlab/index.html}}, the VR tools developed at the Humanities lab of Lund University\footnote{\url{https://www.humlab.lu.se/en/facilities/virtual-reality/}}, and the ``3DMAP-VR'' platform for the data visualization of 3D magneto-hydro-dynamic models of supernova remnants and young stellar objects developed at INAF-Osservatorio Astronomico di Palermo. Here, we present \textit{E0102-VR} \citep{Baracaglia2019}, an experimental VR application designed to explore the scientific potential (and challenges) of VR technology for the visualization and characterization of multi-dimensional datasets in observational astrophysics.

The oxygen-rich supernova remnant (SNR) 1E\,0102.2-7219 \citep[E\,0102 for short;][]{Dopita1981,Tuohy1983} is located in the Small Magellanic Cloud (SMC), at a distance of $\sim$62\,kpc \citep{Graczyk2014,Scowcroft2016}. At optical wavelengths, it displays an intricate set of filaments visible primarily in the line of [O\,\textsc{\smaller III}]\,$\lambda\lambda$4959,5007\AA, but not only \citep[][]{Blair2000, Seitenzahl2018}. These filaments correspond to the outer layers of the progenitor star, which are visible today as they encounter the reverse shock \citep[][]{Sutherland1995}. From their proper motions measured with the \textit{Hubble Space Telescope} (\textit{HST}), \cite{Finkelstein2006} derived an age of (2054$\pm$584)\,yr for the remnant. 

For a SNR in the ``free expansion'' phase of its life \citep{Dopita2003}, interactions with the surrounding medium have not yet significantly altered the chemical composition and/or kinematics of the ejecta knots. If one assumes ballistic expansion for the ejecta in E0102, the depth $z$ of a given knot along the line-of-sight can be trivially derived from its relative radial velocity (with respect to that of the SMC). The actual 3D structure of the oxygen-bright ejecta in E\,0102 can thus be reconstructed from a complete spectro-photometric map of the system \citep{Vogt2010,Vogt2017c}. 

Deep observations of SNR E\,0102 \citep{Vogt2017a} with the \textit{Multi-Unit Spectroscopic Explorer} \citep[MUSE;][]{Bacon2010} integral field spectrograph at the Very Large Telescope revealed a peculiar ring of Ne\,\textsc{\smaller I} and O\,\textsc{\smaller I} emission. This ring, $(0.63\pm0.11)$\,pc in radius, led to the identification (in archival \textit{Chandra X-ray Observatory} data) of the Central Compact Object (CCO) in the system \citep{Vogt2018,Hebbar2019}. The exact nature of the Ne and O optical ring surrounding the CCO remains unclear. \cite{Vogt2018} argue that the excellent spatial alignment between the pc-size optical ring and the CCO, (2054$\pm$584)\,yr after the core-collapse, suggests that the center of the ring corresponds to the supernova explosion site. This location, however, is $\sim$10$^{\prime\prime}\equiv3$\,pc away from the center of symmetry of the very regular X-ray emission of the system \citep{Gaetz2000, Xi2019}, which would then require specific circumstances in the structure of the surrounding medium to be explained. The overall on-sky distribution of the oxygen-rich ejecta visible at optical wavelengths is also significantly more asymmetric than the X-ray emission. The characterization of the actual 3D structure of the optical ejecta in E\,0102 hence appears as an important step towards assembling a coherent evolutionary scenario that could reconcile observations across all wavelength ranges. The VR application \textit{E0102-VR}, that we present in this article, was assembled to that end. The spirit of the application is discussed in more details in Sec.~\ref{sec:app}. With the aid of \textit{E0102-VR}, we characterize the 3D structure of the optical ejecta in SNR E\,0102 in Sec.~\ref{sec:results}. We conclude with a discussion of the potential of VR for the scientific analysis of multi-dimensional datasets in observational astrophysics in Sec.~\ref{sec:summary}. All uniform resource locators (URLs) provided in this article are valid as of November 2019. Wherever available, digital object identifiers (DOI) are quoted instead.

\section{The \textit{E0102-VR} application} \label{sec:app}

\begin{figure}[htb!]
\centerline{\includegraphics[height=6.2cm]{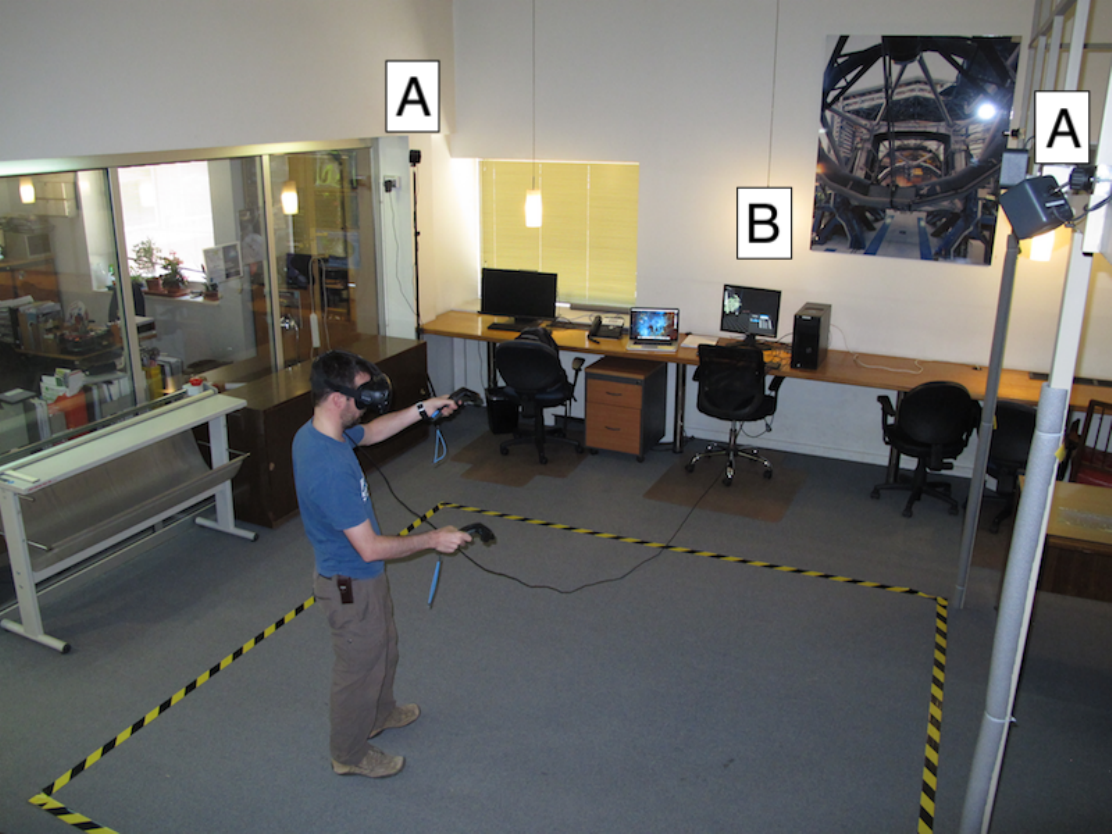}}
\caption{Overview of the VR arena assembled in the library of the Chilean headquarters of ESO in Santiago (Chile), with F.P.A.V in immersion. In addition to the 4\,m$\times$4\,m interaction area, the footprint of the arena includes a tripod supporting one of the two tracking stations (A) and a dedicated computer (B). The VR arena is thus highly flexible and its space can be easily reclaimed for other purposes if/when needed. Safety tape on the floor and foam padding contribute to providing a secure environment for users and by-standers alike.}\label{fig:VRarena}
\end{figure}

\textit{E0102-VR} was developed at the Chilean headquarters of the European Southern Observatory (ESO) in Santiago between June and August 2018. It relied on a 4\,m $\times$ 4\,m ``VR arena'' assembled for this purpose (see Fig.~\ref{fig:VRarena} and Table~\ref{table:char} for details). Two position-tracking sensors (the HTC Vive Base Stations) allow users to move freely over the entire spatial extent of the VR arena, and in so doing, to physically navigate through the virtual world.

\textit{E0102-VR} is an experimental application. The primary goals that drove its development are: a) to explore the capabilities (and limits) of VR to visualize and analyze 3D datasets from observational astrophysics, and b) to identify the challenges and costs associated with assembling such a dedicated VR application. Although this experiment was driven by the specific case of SNR E\,0102, the application itself was developed in such a way that it does not explicitly rely on a specific dataset. Pre-compilation, the dataset at the core of the application can be swapped with no major loss of functionalities. For simplicity and robustness purposes, however, the complied \textit{E0102-VR} application does not provide users with a means to upload different datasets. 

\begin{table}[htb!]
\def\arraystretch{1.5}
\caption{Technical characteristics of \textit{E0102-VR} application and the VR arena used for its development.}\label{table:char}
\vspace{-5pt}
\begin{center}
{\smaller
\begin{tabular}{l p{0.55\columnwidth}}
\hline\hline
\multicolumn{2}{c}{\textit{E0102-VR}}\\
\hline
Development platform & \textsc{unity3d} with custom c\# scripts\\
VR flavor & room-scale\\
Supported headset & HTC Vive\\
Source code license$^{\star}$ & GNU GPL v3\\
\hline
\hline
\multicolumn{2}{c}{VR arena}\\
\hline
Location & Chilean ESO headquarters, Santiago, Chile.\\
Interaction area & 4\,m $\times$ 4\,m\\
VR headset & HTC Vive\\
Support computer & Intel(R) Core(TM) i7-7700K CPU\\[-1.5ex]
& 16GB of RAM\\[-1.5ex]
& NVIDIA GeForce GTX 1060 (6GB)\\[-1.5ex]
& Windows 10 Enterprise\\
\hline
\end{tabular}}
\end{center}
\vspace{-5pt}
{\smaller $\star$: for the source code written by us only. See Sec.~\ref{sec:app} for details.}
\end{table}

\begin{figure}[htb!]
\centerline{\includegraphics[height=6.5cm]{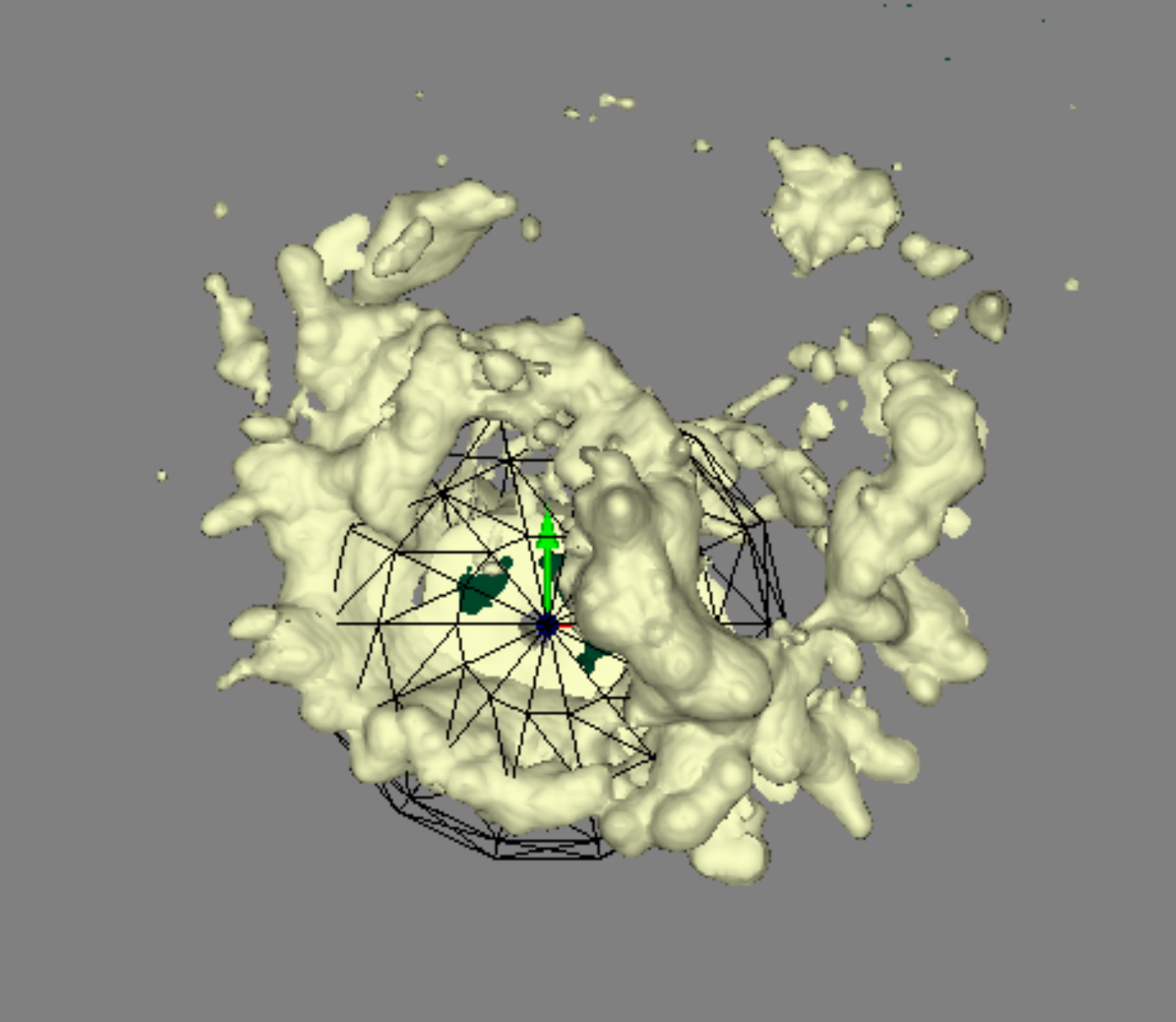}}
\vspace{5pt}
\centerline{\includegraphics[height=6.5cm]{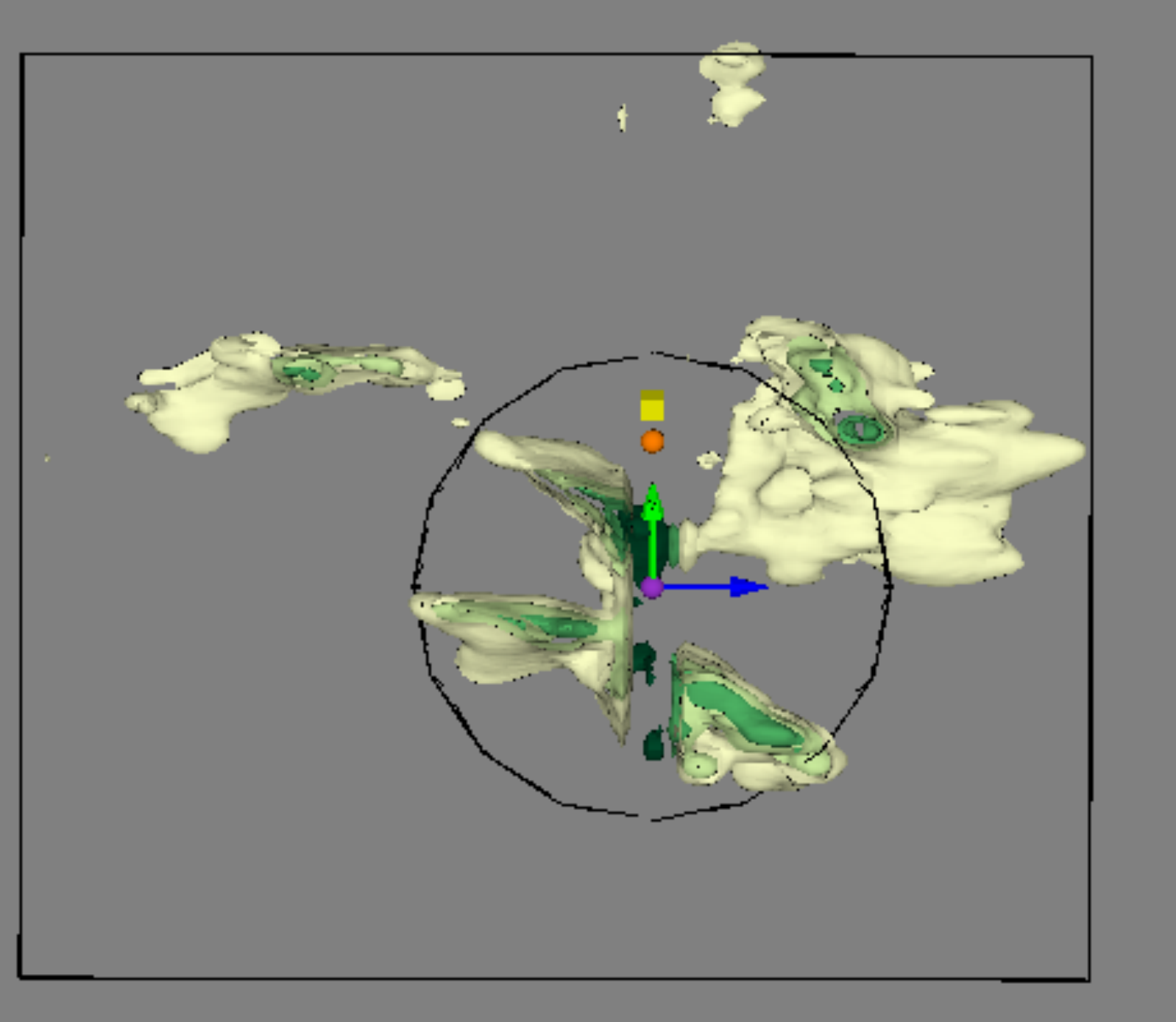}}
\caption{Top: Reconstructed 3D structure of the optical ejecta in SNR E\,0102, as seen from the Earth. The black wireframe sphere is 3\,pc$\equiv$10$^{\prime\prime}$ in radius. North is to the top, and East to the left. Bottom: slice of the oxygen-bright ejecta centered on the CCO (marked as a purple sphere), 1.2\,pc thick with a position-angle of 24$^{\circ}$ West-of-North. Earth is to the right, in the direction of the blue arrow. The North direction (marked by the green arrow) is coming out in front of the slicing plane by 24$^{\circ}$. The funnel structure is clearly visible in this plane, with the CCO at its base. The black circle is 3\,pc$\equiv$10$^{\prime\prime}$ in radius. The center of symmetry of the X-ray emission from SNR E\,0102 (set along the depth axis at the same level as that of the CCO) is marked with a yellow cube. The center of explosion derived by \cite{Finkelstein2006} is marked with an orange sphere. The scale of the model is the same in both panels. The interactive version of this figure, which is is available online at \url{http://fpavogt.github.io/E0102-VR}, allows to rotate/scale the model, alter/remove the clipping planes, and show/hide the 4 intensity layers of [O\,\textsc{\smaller III}] emission corresponding to flux density levels of $[25,12,6,2.5]\times10^{-18}$\,erg\,s$^{-1}$\,cm$^{-2}$\,\AA$^{-1}$.}\label{fig:x3d}
\end{figure}

\textit{E0102-VR} was built using the \textsc{unity3d} software, supplemented with custom \textsc{c\#} scripts. We purposely restricted the scope of the application upfront, as follows. First, the application was developed solely for the HTC Vive: the VR device at our disposal. Whereas this decision undoubtedly restricts the pool of potential users, it allowed us to focus our entire attention on exploring the scientific potential of VR, rather than tackle possible device compatibility challenges which are prone to rapid evolution. Second, the 3D structure of the optical ejecta in SNR E\,0102 is visualized using a set of pre-computed iso-surfaces fed ``as-is'' to the application. Volumetric rendering, an alternative visualization technique to iso-surfaces, is less straightforward to implement, and its potential is already being explored by others \citep[see e.g.][]{Ferrand2016,Ferrand2018}. We also favor iso-surfaces, as they can easily be used, by means of the X3D pathway \citep{Vogt2016}, to generate an interactive 3D model of the ejecta structure publishable in the scientific literature (see Fig.~\ref{fig:x3d}).

The compiled application (which requires an HTC Vive to be run) is freely available online\footnote{at \url{http://fpavogt.github.io/E0102-VR}}. All the source code \textit{written by us} (\href{https://doi.org/10.5281/zenodo.3523259}{DOI: 10.5281/zenodo.3523259}) is released on a dedicated Github repository\footnote{at \url{https://github.com/fpavogt/E0102-VR}} under the GNU General Public License version 3. At the time of publication, the terms of service associated with the use of free assets from the \textsc{unity3d} store (see below) prevent us from releasing the full source code of the application. Readers interested to compile the \textit{E0102-VR} application from source will thus need to download specific supplementary (and free) packages from the \textsc{unity3d} asset store.

The iso-surfaces associated with the [O\,\textsc{\smaller III}] emission in SNR E\,0102 were computed in \textsc{python} using the \textsc{mayavi} package \citep{Ramachandran2011}. We restricted ourselves to a set of 4 iso-surfaces, associated with [O\,\textsc{\smaller III}] emission levels of $[25,12,6,2.5]\times10^{-18}$\,erg\,s$^{-1}$\,cm$^{-2}$\,\AA$^{-1}$. We followed the steps described by \cite{Vogt2017c} to de-blend the two [O\,\textsc{\smaller III}]\,$\lambda\lambda$4969,5007\AA\ lines, and transform the original ``R.A.-Dec.-$\lambda$'' MUSE datacube to an ``$x-y-z$'' 3D map in units of pc\,$\times$\,pc\,$\times$\,pc. The iso-contours generated using \textsc{mayavi} are saved to a single \textsc{obj} file, together with an \textsc{mtl} (material) file specifying the color and transparency of each surface. The choice of this file format was dictated mainly by its compatibility with \textsc{unity3d}, and by the fact that \textsc{obj} files can be easily edited and handled by a wide variety of software. In addition to the 4 iso-surfaces, the final 3D structure fed to the \textit{E0102-VR} application also includes 1) a purple sphere to mark the location of the CCO in the system, which we set as the reference of the coordinate system, and 2) a wireframe sphere of 3\,pc in radius to provide users with an indication of scale. 

\begin{figure*}[htb!]
\centerline{\includegraphics[height=6.5cm]{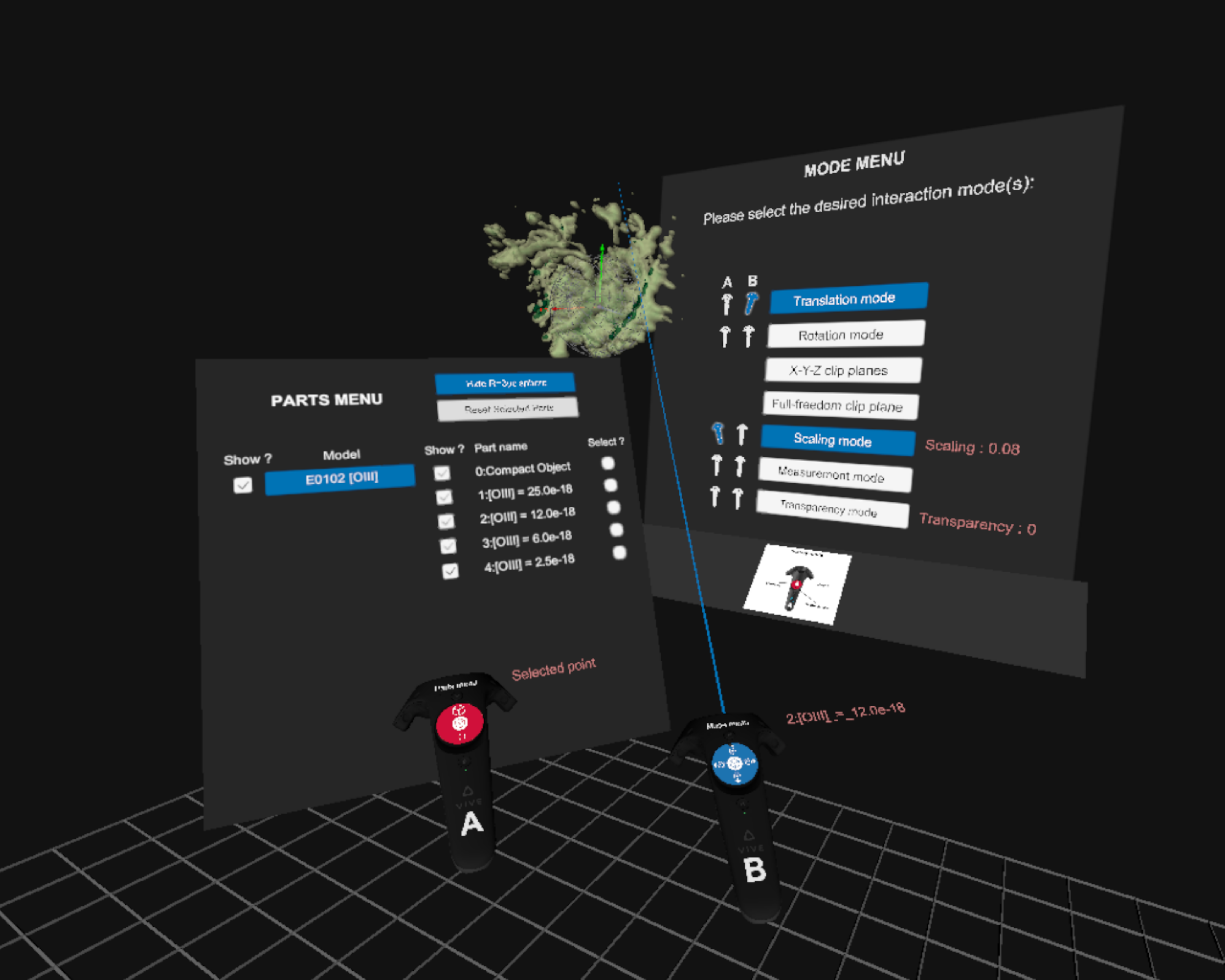}\quad
                  \includegraphics[height=6.5cm]{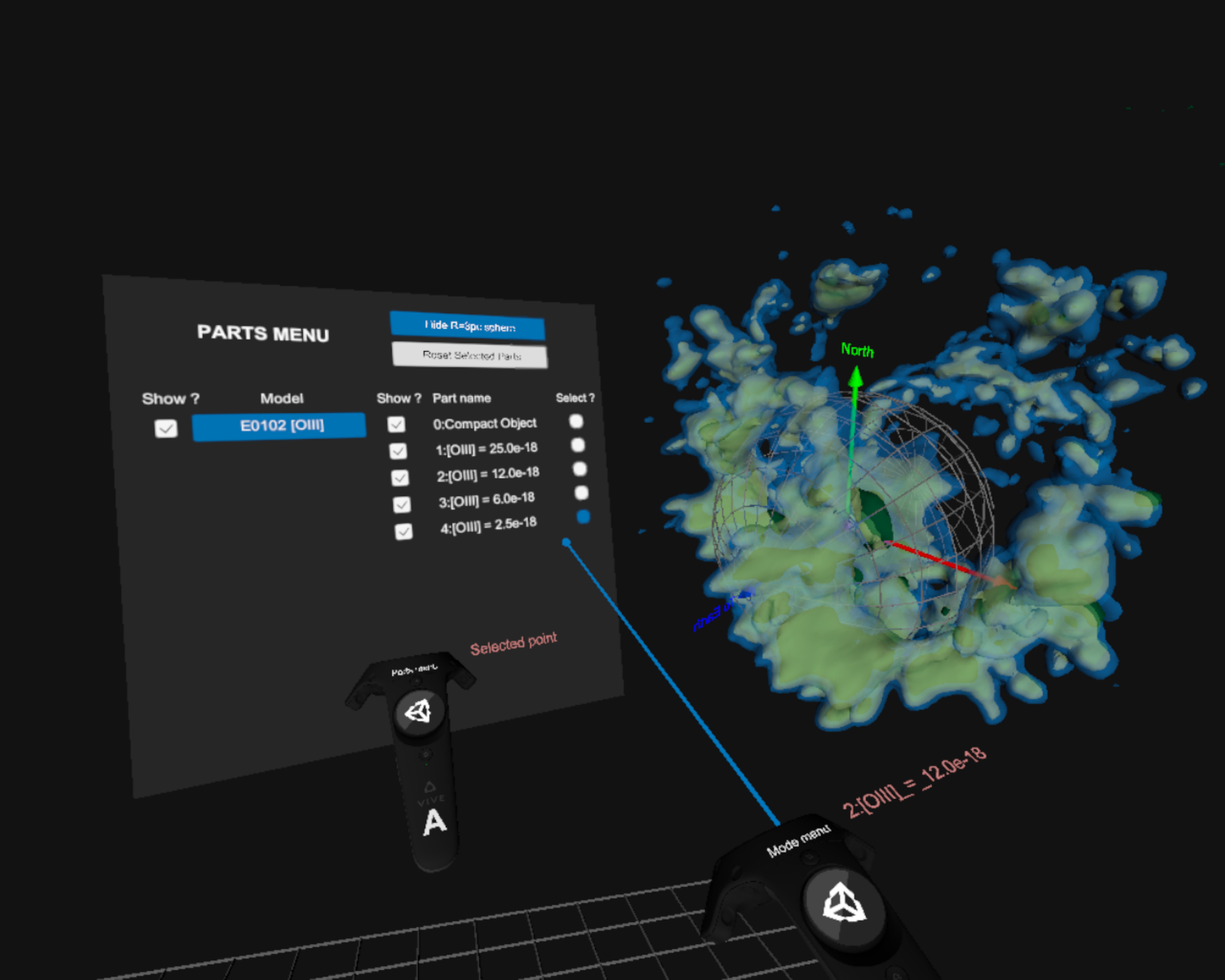}}
\centerline{\includegraphics[height=6.5cm]{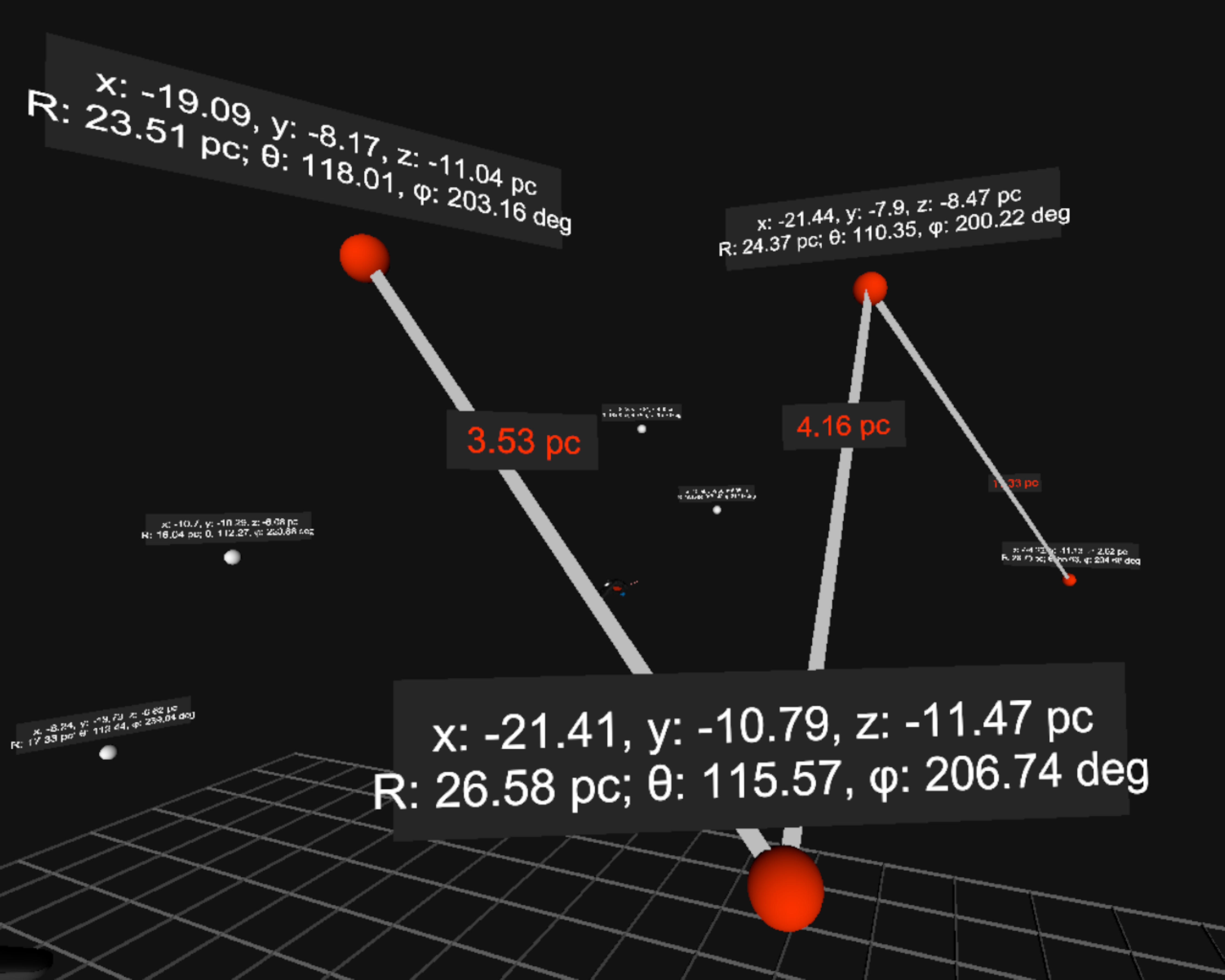}\quad
                  \includegraphics[height=6.5cm]{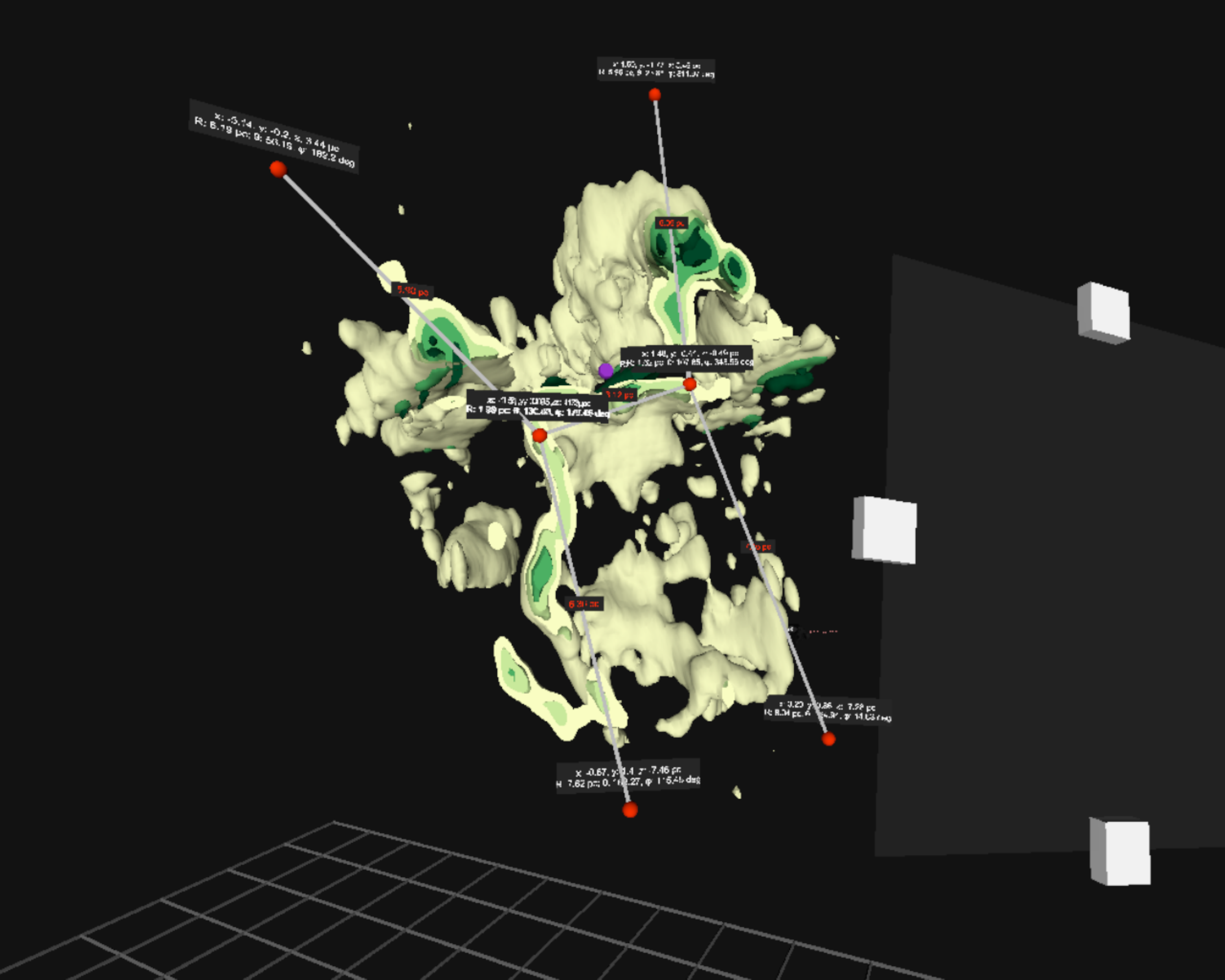}}
\caption{In-app screenshot from \textit{E0102-VR}. Top left: general view. The interaction \textit{mode menu} is visible to the right, while the \textit{parts menu}, attached to the controller ``A'' is visible to the center left. The 3D structure of the optical ejecta in SNR E\,0102 is visible in light green. A blue beam, used to interact with buttons is attached to the controller ``B'', on which the ``translation'' mode is enabled. When users select a specific interaction mode, a panel describing its use (schematically) appears on a dedicated console at the base of the mode menu. Top right: users can use the parts menu (attached to controller ``B'') to select specific model parts, in order to hide/show/recenter them. Bottom left: screenshot demonstrating the ``measurement'' interaction mode. Measurement points can be placed manually by the user in the 3D volume. Selected points (in red) are connected by a straight line with the inter-point distance displayed in the model unit. Bottom right: different interaction modes, assigned to distinct controllers, can be used in parallel. In this example, the ``full freedom clip plane'' (seen to the right) is used together with the ``measurement'' mode to measure the size of different ejecta structures. A demonstration video with in-app footage is available online at \url{https://fpavogt.github.io/E0102-VR/}.}\label{fig:E0102}
\end{figure*}

Within the application, the user primarily interacts with the data by means of a series of distinct and complementary \textit{interaction modes}, which can be enabled or disabled from a dedicated ``mode menu'' (see Fig.~\ref{fig:E0102}). The different interaction modes, some of which can be used in parallel, are: 1) translation, 2) rotation, 3) x-y-z clip planes, 4) full-freedom clip plane, 5) measurement, and 6) transparency. The need to separate the different interaction modes is primarily driven by the limited number of buttons present on the HTC Vive controllers. In that sense, \textit{E0102-VR} is somewhat reminiscent of the \textsc{ds9} software \citep{Joye2003} and the variety of functions that it can assign to the mouse buttons. 

The two clip plane modes, which let the user slice 3D structures to obtain cut views (in the case of \textit{E0102-VR}, either along the X-Y-Z axes, or freely) rely on the \textsc{CrossSectionShader} package\footnote{\url{https://github.com/Dandarawy/Unity3DCrossSectionShader}}. These modes were implemented in an effort to mimic the similar capabilities provided by the X3D pathway \citep{Vogt2017c}. The measurement mode, on the other hand, is a \textit{uniquely-VR} feature, in that it relies on the depth perception provided by the system to the user. Specifically, this interaction mode allows the user to place \textit{measurement points} throughout the 3D environment. Each point displays its cartesian and spherical coordinates. Different measurement points can also be \textit{selected} to measure the distance (in the native model units; pc in the case of SNR E\,0102) between them (see Fig.~\ref{fig:E0102}).

The content of the model, comprised in our case of 6 individual parts, is visible via a dedicated ``parts menu''. This menu takes the form of small panel attached to one of the controller, akin to a painter palette. From this menu, the user can show or hide individual parts of the model, or reset them to their default position (as each part can be grabbed and moved freely).

\begin{figure*}[htb!]
\centerline{\includegraphics[scale=0.15]{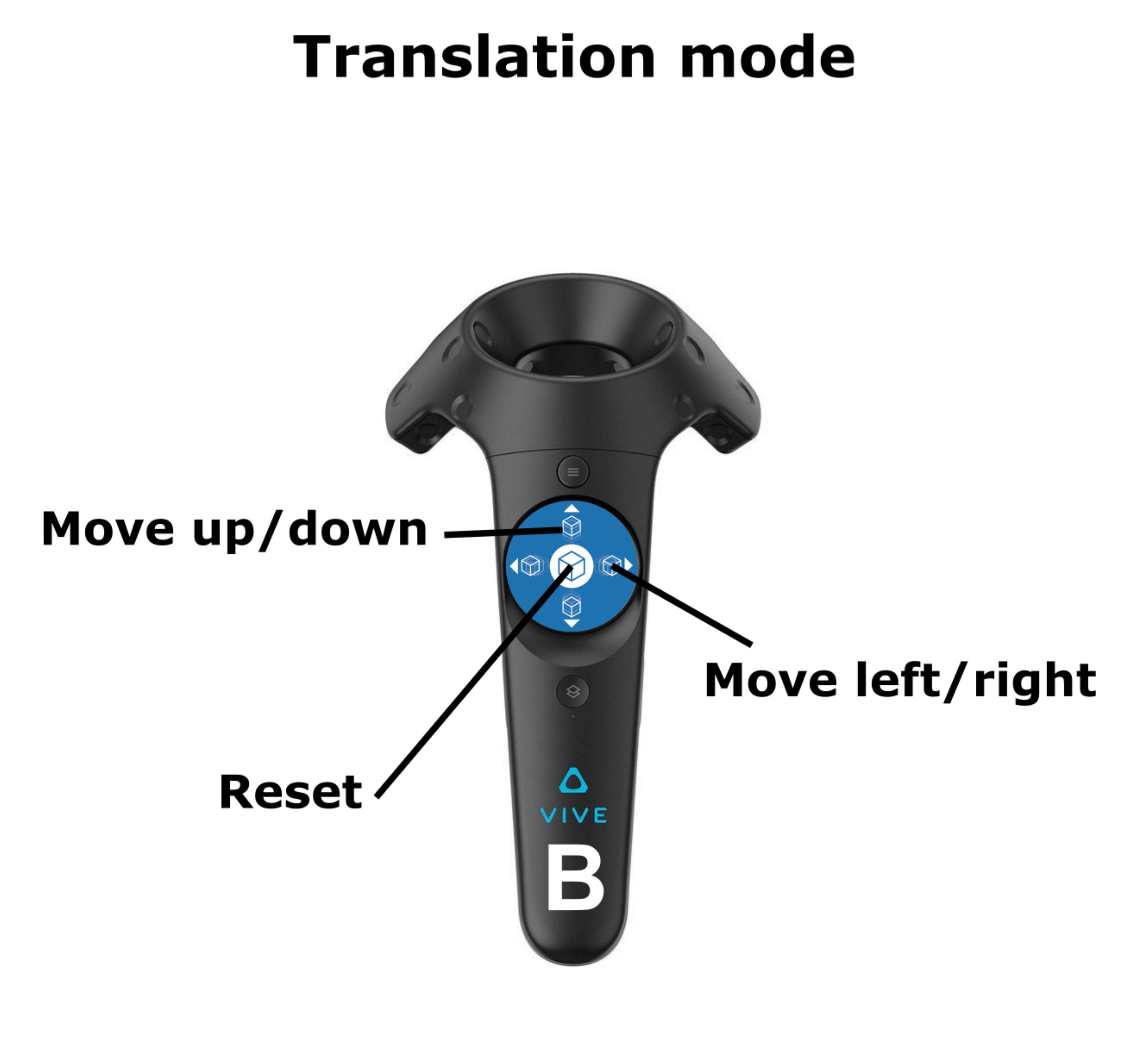}\includegraphics[scale=0.15]{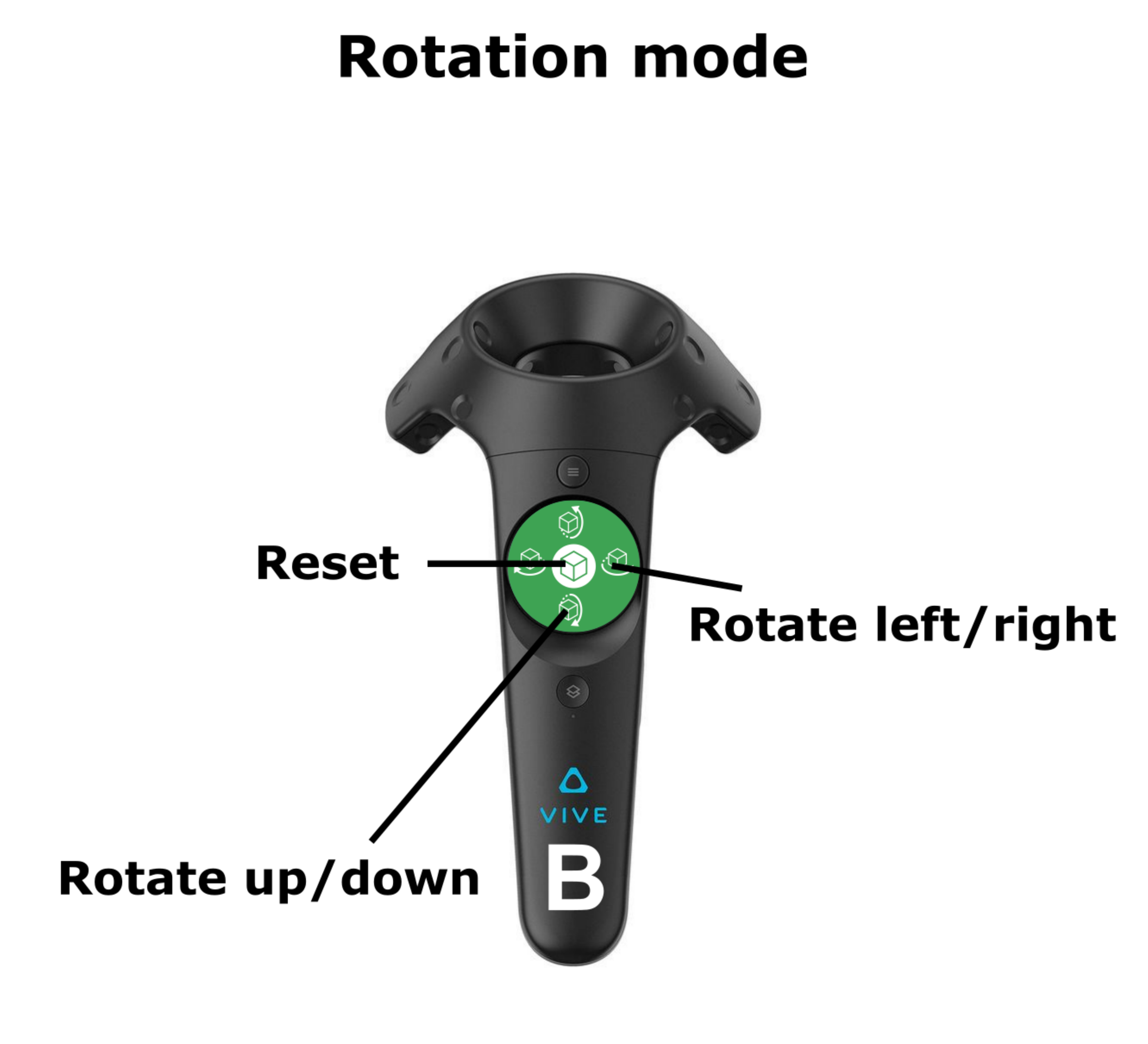}\includegraphics[scale=0.15]{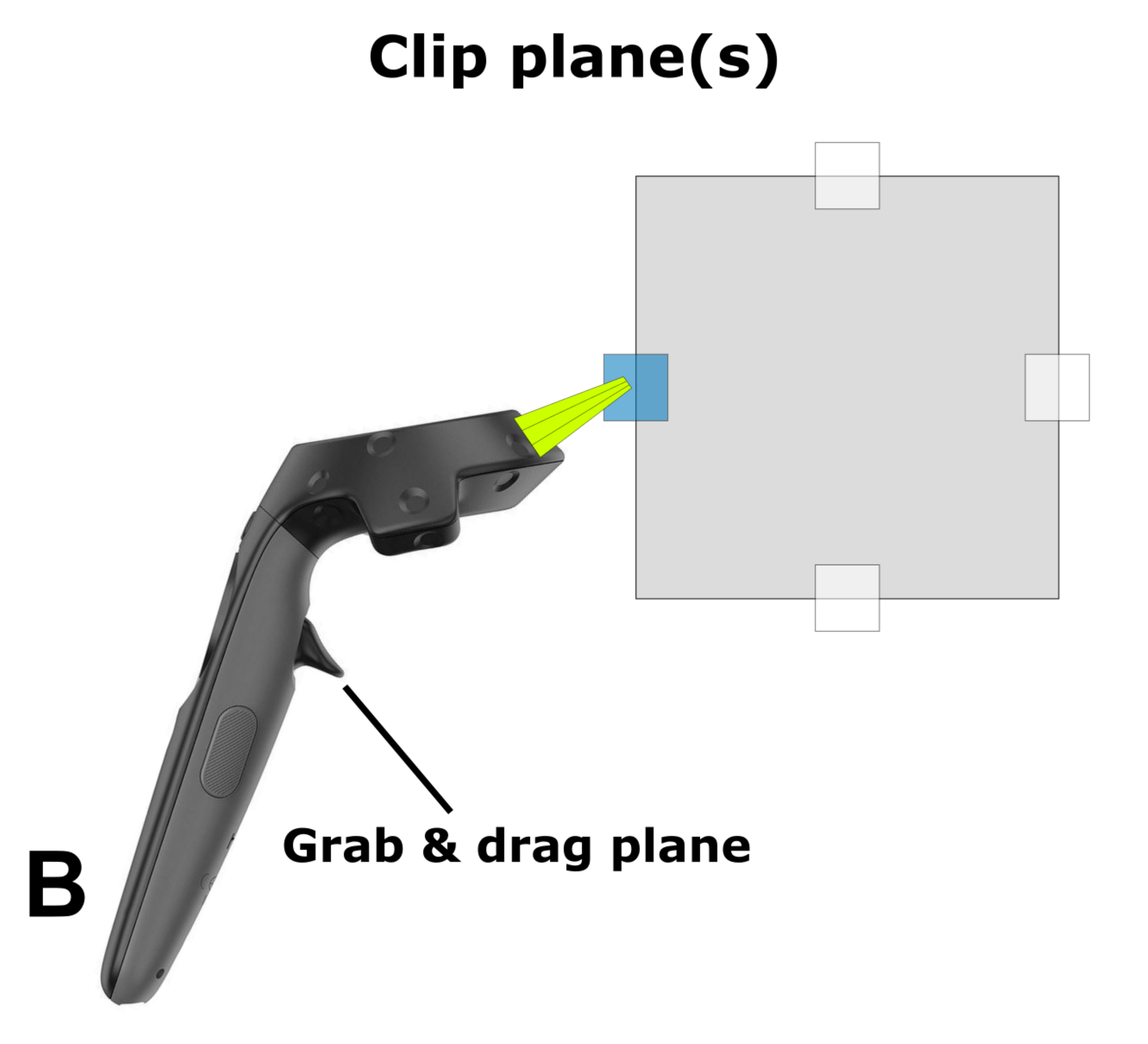}}
\vspace{20pt}
\centerline{\includegraphics[scale=0.15]{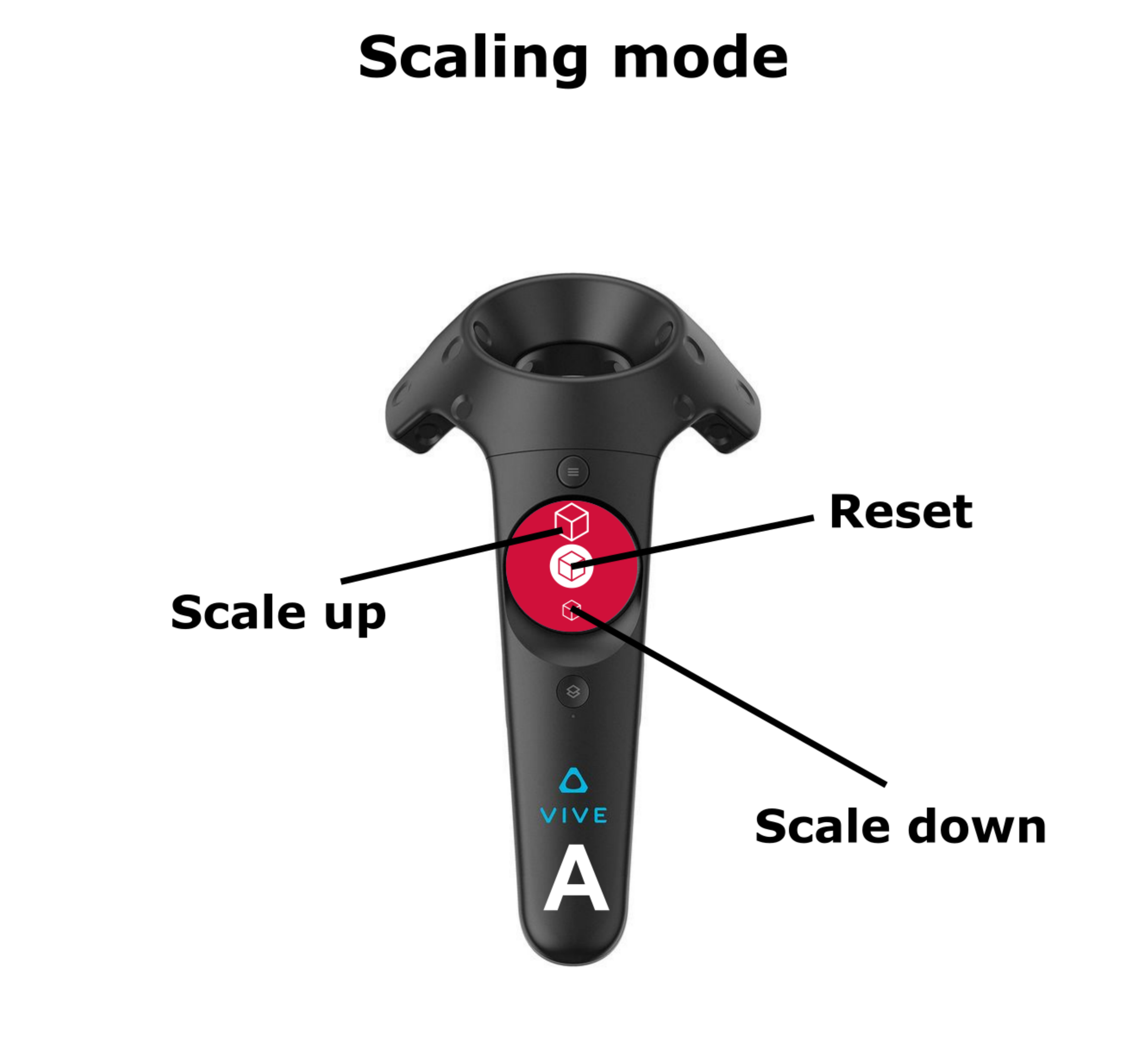}\includegraphics[scale=0.15]{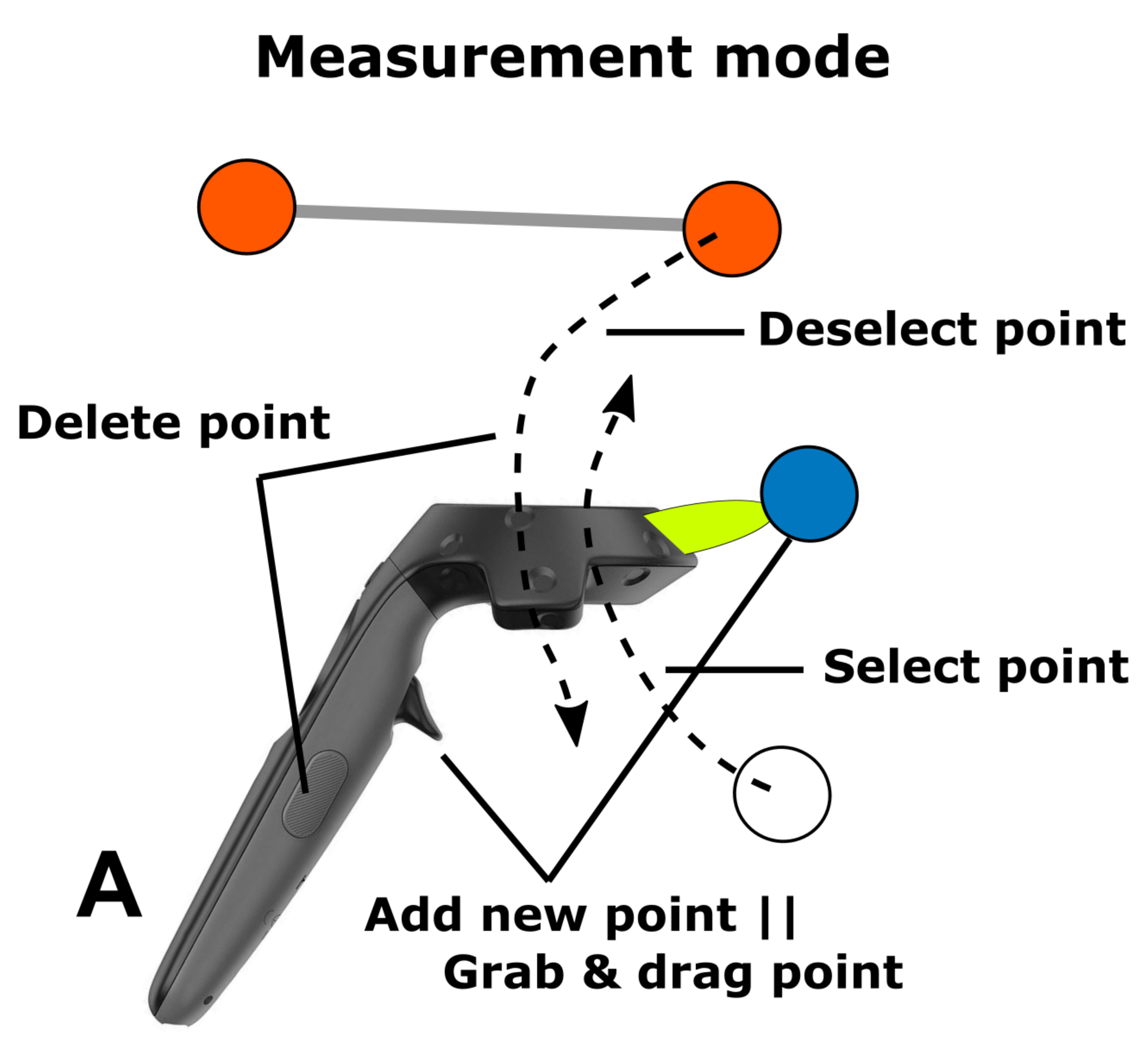}\includegraphics[scale=0.15]{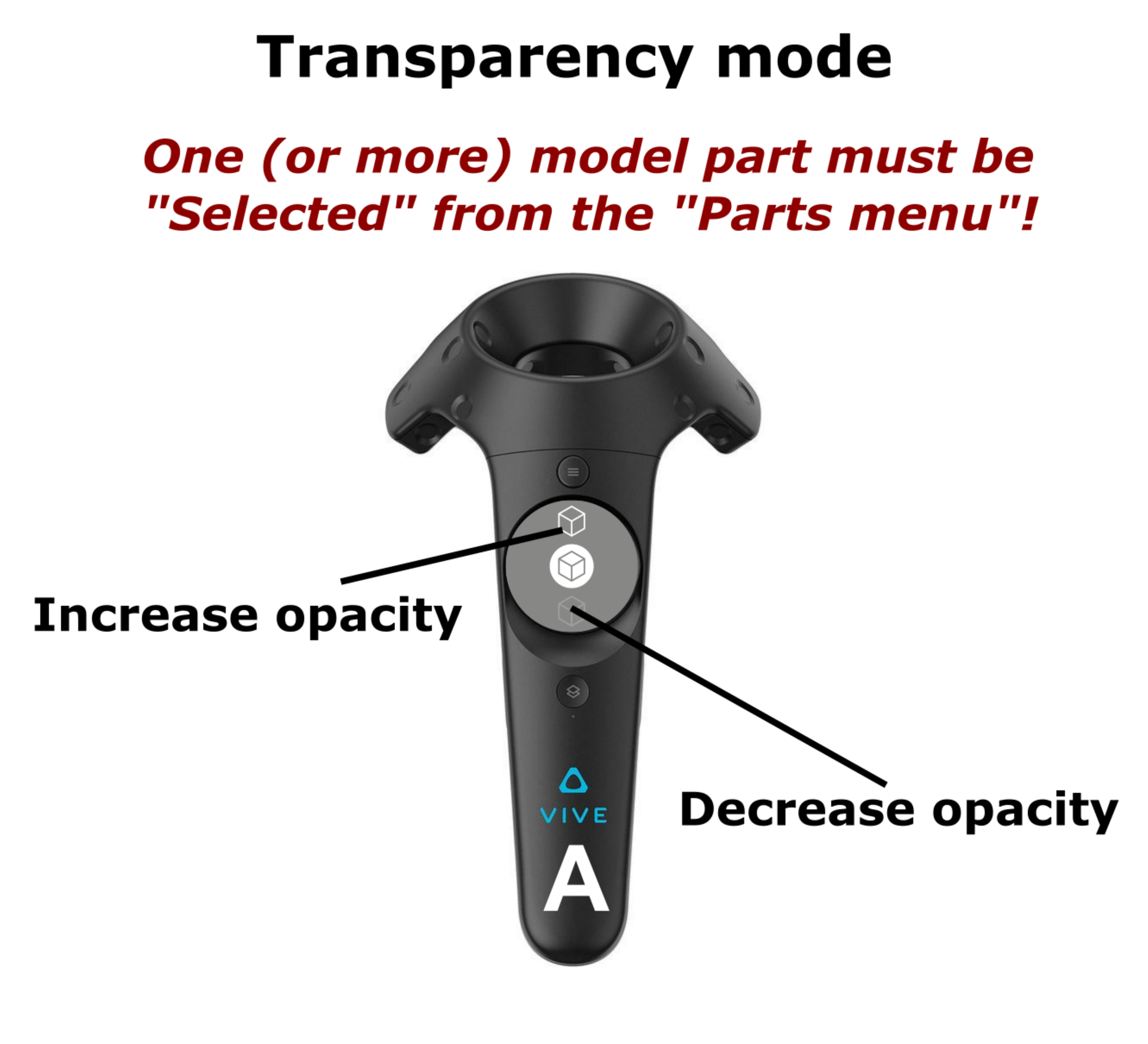}}
\caption{Top: instruction panels used in \textit{E0102-VR} to describe the different interaction modes associated to the controller ``B''. These were implemented in lieu of a formal user manual, which is instead available online on the dedicated app website. Each instruction panel is visible immediately below the mode menu when the associated interaction mode is enabled by the user (see Fig.~\ref{fig:E0102}). Bottom: idem, but for the interaction modes associated to the controller ``A".}\label{fig:pictograms}
\end{figure*}

We have made some efforts to ensure that our choice of buttons associated with the different interaction modes is reasonably comfortable and intuitive. We opted against the inclusion of a written instruction manual inside \textit{E0102-VR}. Whereas such a manual is present on the app website\footnote{\url{https://fpavogt.github.io/E0102-VR/tutorial.html}}, we favored visual cues inside the app to aid the users in understanding the different interaction modes. The controller buttons carry explicit symbols and/or simple words (some of which change depending on the chosen interaction mode) illustrating the associated function. Descriptive panels are also visible (immediately below the mode menu) when a specific interaction mode is enabled (see Figs.~\ref{fig:E0102} and ~\ref{fig:pictograms}). Undeniably, however, \textit{E0102-VR} remains an experimental application with room for improvement on the user-interaction front. In particular, the $\beta$-testing of the \textit{E0102-VR} application by members of the ESO staff, Fellow, and student communities revealed that the more advanced interaction modes that rely on the active interaction of the users with the model (such as the measurement mode, for example) are not immediately understood by all users. We note, however, that all users appeared to feel rapidly comfortable with even the most complex of the interaction modes once understood.

\section{The optical ejecta in E\,0102} \label{sec:results}
The 3D structure of the oxygen-bright ejecta in SNR E\,0102 was reconstructed by \cite{Vogt2010}, based on observations from the Wide-Field Spectrograph \cite[WiFeS;][]{Dopita2007,Dopita2010}. The structure was published as an interactive \textsc{PDF} file, with the significant drawback that it forced readers to use specific software to access it. \cite{Vogt2017c} revisited the problem using a new set of WiFeS observations and the X3D pathway \citep{Vogt2016}.

The 3D map of the oxygen-bright ejecta in SNR E\,0102 discussed here and included in the \textit{E0102-VR} application is based on MUSE observations acquired under programme 297.D-5058[A] (P.I. Vogt) and described in detail by \cite{Vogt2017a}. Spatially, this map has a seeing-limited resolution of 0.7$^{\prime\prime}\equiv$0.2\,pc, which is $\sim$2 times sharper than the previous WiFeS observations. Along the line-of-sight, however, the spectral resolution of MUSE (and thus the spatial resolution of the derived map) at the wavelength of [O\,\textsc{\smaller III}]\,5007\AA\ is $\sim$2.3 times worse than achieved with WiFeS in its high resolution mode. Whereas the MUSE 3D map can be best visualized and characterized using the \textit{E0102-VR} application, we also make it accessible as an online interactive figure (available at \url{http://fpavogt.github.io/E0102-VR}). With it, any reader ought to be able identify the different structures present within the ejecta (and discussed below) without the need for an HTC Vive.

The spatial distribution of the oxygen-bright knots of ejecta in SNR E\,0102 is complex, but not disorganized. Large, irregular walls and sheets of emission, up to a few pc in length, separate equally large voids (in the optical line of [O\,\textsc{\smaller III}]). The walls, spatially unresolved in the MUSE datacube, have a thickness $<0.5$\,pc. From the \textit{HST} observations of the system \citep{Finkelstein2006}, their thickness is likely $\lesssim0.05$\,pc. This is reminiscent of the equally-structured ejecta in Cassiopeia A \citep{Isensee2010,DeLaney2010,Milisavljevic2013}, which may be tracing the boundaries of plumes of $^{56}$Ni-rich ejecta \citep{Blondin2001,Milisavljevic2015,Wongwathanarat2017}.

On the far-side of the remnant (i.e.\ the side facing away from Earth), a large funnel-like structure is particularly noticeable. The structure is $(7.0\pm0.5)$\,pc long with a cross-section of $(4.0\pm0.5)\times(3.0\pm0.5)$\,pc$\times$pc at half-maximum, and does not contain any [O\,\textsc{\smaller III}] emission at a level $\geq2.5\times10^{-18}$ erg\,s$^{-1}$\,cm$^{-2}$\,\AA$^{-1}$ (see Fig.~\ref{fig:x3d}). Along the $z\equiv$ radial velocity direction, the base of the funnel is coincident with the rest-frame of the SMC to within the spectral resolution of MUSE at a 1$\sigma$ level. Spatially, the base of the funnel has a size of  $(3.0\pm0.5)\times(1.5\pm0.5)$\,pc$\times$pc with the long axis rotated by $(40\pm10)^{\circ}$ East-of-North, and seem to coincide with the location of the Ne-O ring surrounding the CCO. On the near-side of the remnant, an inverted cone-like opening, $(5.0\pm0.5)$\,pc in diameter and $(3\pm0.5)$\,pc in length, could be described as an (asymmetric, shallower) counterpart funnel. 

One may be tempted to interpret the apparent alignment between the base of the funnel structure and the Ne-O ring as another suggestion that this location was the actual site of the SN explosion in SNR E0102. We do not venture into this discussion here, which will very strongly benefit from a refined measurement of the center of explosion (Banovetz et al., in preparation) based on the ejecta's proper motion measured over a longer time-span than accessible to \cite{Finkelstein2006}. 

\section{VR lessons learned from \textit{E0102-VR}}\label{sec:summary}

The depth perception provided to the user by the \textit{E0102-VR} application greatly helps in understanding the complex 3D structure of the oxygen-bright ejecta in SNR E\,0102. Yet, this fact alone does not suffice to justify the efforts required to assemble the application in the first place: for example, the interactive 3D model of the ejecta assembled using the X3D pathway can also provide the user with an excellent understanding of the 3D structure with significantly less efforts, and no specific gear required. What makes the \textit{E0102-VR} application stand-out from a scientific perspective, however, is that it provides a rapid, easy, and precise means to measure distances and angles in 3D. Characterizing the size of structures in SNR E\,0102 only requires a few seconds in VR, as measurement points can be accurately positioned in 3D space without any ambiguity: a feat which is nigh-impossible in 2D figures, be they interactive or not. From that perspective, we foresee that the ability to interact with the 3D model, and thus the use of hand controllers, will be an essential component of any VR application aimed at the scientific analysis of astrophysical datasets. 

The benefits of a room-scale component, allowing the user to physically ``move around'' in virtual space, are less evident at this stage. During the $\beta$-testing of the \textit{E0102-VR} application, we noticed that users tend to rotate and translate the model, rather than walk around it. This behavior may be a consequence of the disruptive presence of cables connecting the HTC Vive headset to the desktop computer, which tend to hinder user-motion. The removal of these cables (in favor of a wireless connection) that took place in June 2019 may possibly lead to an increased number of ``wandering users'' in the VR arena. 

Evidently, the mere fact of measuring 3D distances and angles within seconds is still very unlikely to justify (in most cases) a time investment of 1 full-time-employee over 3-months to assemble a dedicated \textit{E0102-VR}-like application. Furthermore, as of 2019, VR technology relies on a series of programming languages and software entirely foreign to the vast majority of observational astronomers. We are thus convinced that the future of VR in our field does not reside with dedicated, one-off applications like \textit{E0102-VR}. Rather, we see the wide-spread use of VR in observational astrophysics (for scientific analysis purposes) tied to the development of a high-level, generic application dedicated to basic 3D data inspection and characterization tasks: a type of application epitomized by \textsc{ds9} for 2D datasets. The development of specific, one-off VR applications certainly already lies well within the reach of motivated teams and individuals, but it is highly unlikely for VR to become a wide-spread scientific tool in astrophysics without the existence of a robust, polyvalent, and generic application first.

By design, none of the functionalities of the \textit{E0102-VR} application are intrinsically tied to the underlying 3D model, to the extent that the model could be swapped (in \textsc{unity3d}) with another without any major loss of functionality. Other experiments, including the VRlab\footnote{\url{http://www.usm.uni-muenchen.de/vrlab/index.html}}, have also started to explore the enticing prospect of generic loading modules for astrophysical datasets. This suggests that a ``ds9-VR'' --an application dedicated to providing observational astronomers generic visualization, exploration and measurement tools for 3D datasets-- is both feasible and technically viable, given existing VR hardware and software. With the \textit{E0102-VR} application and its measurement mode, we have demonstrated some of the unique scientific potential that such a tool could offer. Other useful VR interaction modes include, for example, the measurement of volumes, the ability to draw and annotate models in 3D, and multi-user interactions. Different means of visualizing data, be it with the live reconstruction of iso-contours or full volumetric rendering, also appears highly desirable.

The ongoing democratization of room-scale, plug-and-play VR devices will likely only prompt a limited number of curious individuals and institutions to explore the potential of VR technology for observational astronomy in the months and years ahead. It is only if these pioneering efforts can be merged into a robust, high-level tool that we would foresee the rest of the community to follow suit, and start exploiting the potential of this technology for scientific purposes.


\section*{Acknowledgments}
{\smaller 
We thank Joe Anderson, James Leftley, Chiara Mazzucchelli, B\'arbara N\'u\~nez, \'Alvaro Ribas, Eleonora Sani, Alejandro Santamaria Miranda, Fernando Selman, and Romain Thomas for $\beta$-testing the \textit{E0102-VR} application. 

This research has made use of the following \textsc{python} packages: \textsc{aplpy} \citep{Robitaille2012}, \textsc{astropy} \citep{AstropyCollaboration2013,AstropyCollaboration2018}, \textsc{brutifus} \citep{Vogt2019}, \textsc{matplotlib} \citep{Hunter2007}, \textsc{mayavi} \citep{Ramachandran2011}, and \textsc{statsmodel} \citep{Seabold2010}. This research has also made use of Aladin \citep{Bonnarel2000}, of \textsc{saoimage ds9} \citep{Joye2003}, of NASA's Astrophysics Data System, and of the NASA/IPAC Extragalactic Database \citep[NED;][]{Helou1991} which is operated by the Jet Propulsion Laboratory, California Institute of Technology, under contract with the National Aeronautics and Space Administration. This work has made use of data from the European Space Agency (ESA) mission {\it Gaia} (\url{https://www.cosmos.esa.int/gaia}), processed by the {\it Gaia} Data Processing and Analysis Consortium (DPAC, \url{https://www.cosmos.esa.int/web/gaia/dpac/consortium}). Funding for the DPAC has been provided by national institutions, in particular the institutions participating in the {\it Gaia} Multilateral Agreement. Specifically, this work relied on data from the {\it Gaia} \citep{GaiaCollaboration2016} Data Release 2 \citep{GaiaCollaboration2018} to refine the World-Coordinate-System (WCS) solution of a MUSE datacube.

This work was supported by Science Support Discretionary Funds 18/32 C, 18/33 C, and 19/01 from the European Southern Observatory. Based on observations made with ESO Telescopes at the La Silla Paranal Observatory under programme ID 297.D-5058[A].}

\bibliographystyle{aa}
\bibliography{bibliography_fixed.bib}

\end{document}